# Influence of core *NA* on Thermal-Induced Mode Instabilities in High Power Fiber Amplifiers

Rumao Tao, Pengfei Ma, Xiaolin Wang*, Pu Zhou**, Zejin Liu

*College of Optoelectric Science and Engineering, National University of Defense Technology,
Changsha, Hunan 410073, China*
e-mail: *chinawxllin@163.com; **zhoupu203@163.com

*Abstract*—We report on the influence of core *NA* on thermal-induced mode instabilities (MI) in high power fiber amplifiers. Influence of core *NA* and *V*-parameter on MI has been investigated numerically. It shows that core *NA* has larger influence on MI for fibers with smaller core-cladding-ratio, and the influence of core *NA* on threshold is more obvious when the amplifiers are pumped at 915nm. The dependence of threshold on *V*-parameter revealed that the threshold increases linearly as *V*-parameter decreases when *V*-parameter is larger than 3.5, and the threshold shows exponentially increase as *V*-parameter decreases when *V*-parameter is less than 3.5. We also discussed the effect of linewidth on MI, which indicates that the influence of linewidth can be neglected for linewidth smaller than 1nm when the fiber core *NA* is smaller than 0.07 and fiber length is shorter than 20m. Fiber amplifiers with different core *NA* were experimentally analyzed, which agreed with the theoretical predictions.

*Index Terms*—Fiber amplifier, core *NA*, mode instabilities, thermal effects

## I. Introduction

It is desirable to have high power fiber laser systems with diffraction limited beam quality, which are attractive sources for many applications, such as coherent lidar system, nonlinear frequency conversion, coherent beam combining [1-3]. Generally, large mode area (LMA) fibers are employed to mitigate nonlinear effects and enable higher power scaling [4], which inevitably results in that the fiber supports the propagation of a few modes and the onset of new phenomenon- thermal-induced mode instabilities (MI) [5, 6]. The onset of MI degrades the beam quality and currently limits the further power scaling of ytterbium doped fiber laser systems with diffraction-limited beam quality. Much work on MI has been carried out [7-22] and influence of various fiber parameters on MI, such as core size [14, 17, 22], pump cladding size [18, 19, 20], dopant area [14, 18, 19, 21], has already been studied theoretically and experimentally to achieve further insight of MI. *V*-parameter, determined by core size and core *NA*, is an important character of fiber laser, which determines the number of support mode in core and the constraining capability of the core on fiber mode. Although the influence of *V*-parameter on MI with core *NA* fixed and core size varying has already been investigated [17, 22], little work on the influence of *V*-parameter has been carried out from the aspect of core *NA* [14], which is different from the former case in the effect of gain saturation.

In this paper, the influence of core *NA* on MI in various fibers has been investigated numerically, which has also been studied simultaneously from the aspect of *V*-parameter. The influence of linewidth on MI has also been discussed theoretically. More importantly, a high power master oscillator power amplifier has been setup, which enables us to compare experimental results to the theoretical results. Agreement between theoretical predication and experimental results has been achieved.

## II. Theoretical Study

In theoretical study, the fraction of high order mode (HOM) in signal laser power is used to define the threshold of MI [15, 23]. For the case that MI is seeded by intensity noise of the signal laser, the fraction of high order mode can be expressed as [20, 22]

$$\xi(L) \approx \xi_0 \exp\left[\int_0^L dz \iint g(r,\phi,z)(\psi_2\psi_2 - \psi_1\psi_1) r dr d\phi\right] \quad (1)$$
$$+ \frac{\xi_0}{4}\sqrt{\frac{2\pi}{\int_0^L P_1(z)|\chi''(\Omega_0)|dz}} R_N(\Omega_0) \exp\left[\int_0^L dz \iint g(r,\phi,z)(\psi_2\psi_2 - \psi_1\psi_1) r dr d\phi + \int_0^L P_1(z)\chi(\Omega_0)dz\right]$$

with

$$\chi(\Omega) = 2\frac{n_0\omega_2^2}{c^2\beta_2}\text{Im}\left(4n_0\varepsilon_0 c \iint \bar{h}_{12}\psi_1\psi_2 r dr d\phi\right) \quad (2a)$$

$$\bar{h}_{kl}(r,\phi,z) = \frac{\eta}{\pi\rho C}\left(\frac{v_p - v_s}{v_s}\right)\sum_v \sum_{m=1}^{\infty} \frac{R_v(\delta_m,r)}{N(\delta_m)}\frac{B_{kl}(\phi,z)}{\alpha\delta_m^2 - j\Omega} \quad (2b)$$

$$N(\delta_m) = \int_0^R r R_v^2(\delta_m, r) dr \qquad (2c)$$

$$B_{kl}(\phi, z) = \int_0^{2\pi} d\phi' \int_0^R g_0 R_v(\delta_m, r') \cos v(\phi - \phi') \frac{\psi_k(r', \phi') \psi_l(r', \phi')}{(1 + I_0 / I_{saturation})^2} dr', \quad k \neq l \qquad (2d)$$

where $\eta$ is the thermal-optic coefficient, $\rho$ is the density, $C$ is the specific heat capacity, and $R$ is the radius of the inner cladding, $g(r, \phi, z)$ is the gain distribution in fiber and $\psi_2(r,\phi)$ and $\beta_2$ is the normalized mode profiles and propagation constant of HOM (LP$_{11}$ in the paper). $g_0$ and $I_{saturation}$ is the small signal gain and saturation intensity, respectively. $R_v(\delta_m, r) = J_v(\delta_m r)$ ($J_v$ represents Bessel functions of the first kind) and $\delta_m$ is the positive roots of $\delta_m J_v'(\delta_m R) + h_q / \kappa J_v(\delta_m R) = 0$ ($h_q$ is the convection coefficient for the cooling fluid and $\kappa$ is the thermal conductivity). $R_N(\Omega)$ is the relative intensity noise of the input signal, $\xi_0$ is the initial HOM content. The MI threshold is defined as the pump power at which the fraction of HOM in output power is 0.05.

MI threshold as a function of core *NA* has been calculated in Fig. 1, where 20/400 denotes fiber with core/cladding diameter being 20μm/400μm. The parameters used in the calculation are listed in table I. It is shown in Fig. 1 that threshold power increases with decrease of core *NA*. For the case that core *NA* decrease from 0.07 to 0.045, the threshold power increases by 57%, 25%, 16% and 11% for 20/400, 25/400, 30/400, 30/250 fiber, respectively, when the amplifiers are pumped at 976nm. The threshold power increase is 22% for 30/250 fiber when pumped at 915nm, which is larger than that achieved by pumping at 976nm. We can conclude that, as the core *NA* decreasing, the threshold power increases more obviously for fiber amplifiers with smaller core-cladding-ratio or pumped at 915nm other than 976nm.

TABLE I
PARAMETERS OF TEST AMPLIFIER

| | |
|---|---|
| $n_{clad}$ | 1.45 |
| $\lambda_p$ | 976nm/915nm |
| $\lambda_s$ | 1064nm |
| $h_q$ | 5000 W/(m$^2$K) |
| $\eta$ | 1.2×10$^{-5}$ K$^{-1}$ |
| $\kappa$ | 1.38 W/(Km) |
| $\rho C$ | 1.54×10$^6$ J/(Km$^3$) |
| $\xi_0$ | 0.01 |
| $R_N(\Omega)$ | 1×10$^{-10}$ |
| $P_0$ | 90W |
| $\sigma_s^a$ | 6.00×10$^{-27}$ m$^2$ |
| $\sigma_s^e$ | 3.58×10$^{-25}$ m$^2$ |
| $\sigma_p^a$ | 2.47×10$^{-24}$ m$^2$ |
| $\sigma_p^e$ | 2.44×10$^{-24}$ m$^2$ |

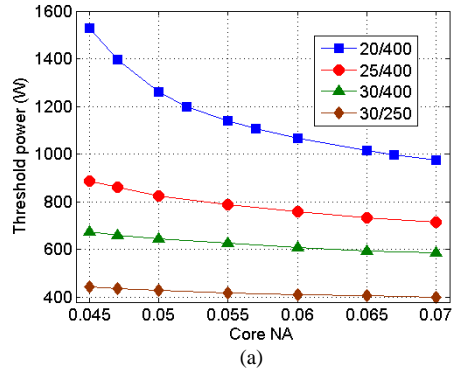

(a)

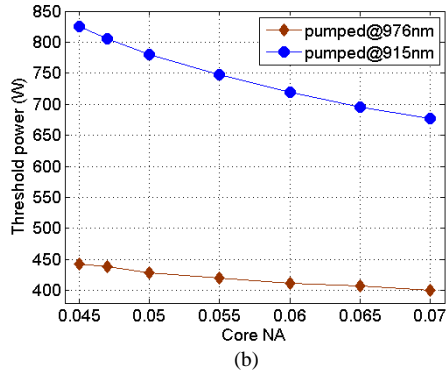

(b)

Fig. 1 Threshold as a function of core *NA*.

To further study the influence of core *NA* on MI, Fig. 2 plotted the mode profile of LP$_{01}$ and LP$_{11}$ mode with different core *NA* or *V*-parameter. It reveals that LP$_{11}$ mode penetrated deeper in the cladding area [24]. By introducing the overlap integral between

mode profile with doped core area

$$\Re = \int_0^{2\pi} \int_0^{R_{core}} \psi_{mn}(r,\phi)\psi_{mn}(r,\phi) r dr d\phi / \iint \psi_{mn}(r,\phi)\psi_{mn}(r,\phi) r dr d\phi \quad (3)$$

the influence of NA on overlap can be analyzed quantitatively, which is presented in Table II. As core NA and V-parameter decreases, $LP_{11}$ mode penetrated deeper while the penetration increment of $LP_{01}$ mode is ignorable, which results in the decrease of overlap between dopant area and $LP_{11}$ mode is larger than that between dopant area and $LP_{01}$ mode. These phenomena are similar to the partial doping in fiber core, and consequently increase the MI threshold power. It can be seen that the expanding of $LP_{11}$ mode into the clad for 30/400 fiber is relatively small compared with that in 20/400 fiber during the same change range of the core NA, which is due to larger value of V-parameter and results that the MI threshold enhancement is smaller for 30/400 fiber as shown in Fig. 1. It also revealed from Fig. 2 that the constraining capability of fiber core on the $LP_{11}$ mode is significantly weakened when V-parameter is 2.66, and most of the mode power is contained in the core when V-parameter is larger than 3.54.

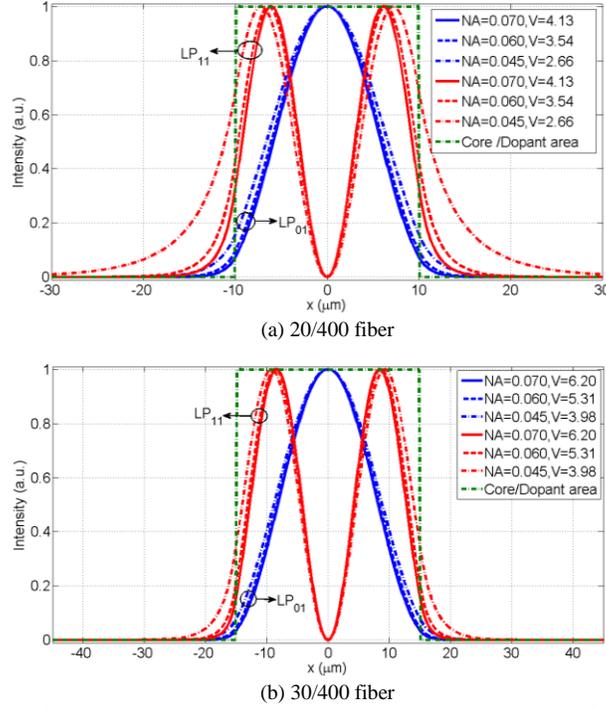

(a) 20/400 fiber

(b) 30/400 fiber

Fig. 2 The mode profile of $LP_{01}$ and $LP_{11}$ in fiber with different core NA.

TABLE II
OVERLAP INTEGRAL OF MODE PROFILE AND DOPANT AREA

| Fiber type | Mode | NA=0.045 | NA=0.06 | NA=0.07 |
|---|---|---|---|---|
| 20/400 | $LP_{01}$ | 0.8629 | 0.9314 | 0.9534 |
|  | $LP_{11}$ | 0.4979 | 0.7882 | 0.8632 |
| 30/400 | $LP_{01}$ | 0.9182 | 0.9755 | 0.9837 |
|  | $LP_{11}$ | 0.7833 | 0.9319 | 0.9557 |

To investigate the influence of V-parameter on MI, the threshold as a function of V-parameter is presented in Fig. 3. It can be seen that the threshold increases linearly as V-parameter decreases when V-parameter is larger than 3.5. When V-parameter is less than 3.5, the threshold shows exponentially increase as V-parameter decreases, which is due to that the constraining capability of the fiber on $LP_{11}$ mode weaken significantly as shown in Fig. 2. When the V-parameter is equal, the threshold is larger for fiber with smaller core size, which is due to the difference of gain saturation [19].

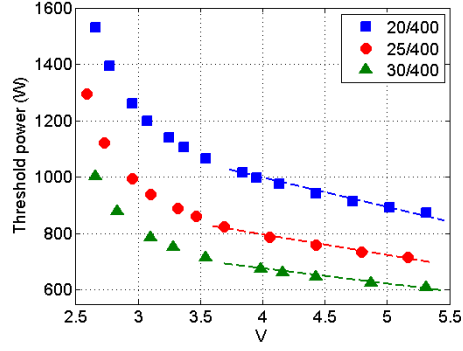

Fig. 3 Threshold as a function of *V* parameter for different types of fiber.

In [20, 22], the linewidth of the signal laser has not been considered, which may play a role in MI [25-27]. Theoretical study shows that the linewidth of the signal laser has negligible effect on MI when the two interfering fields are of time synchronization or the mode walk-off time on the gain fiber length *L* was far less than the signal coherence time ($(1/v_0 - 1/v_1)L \ll \Delta v_s^{-1}$, where $v_{0,1}$ is the mode speed and $\Delta v_s$ is the linewidth of the signal laser) [25, 26]. Define the maximal linewidth that linewidth has negligible effect as $\Delta v_0$, we have

$$\Delta v_0^{-1} = (1/10)(1/v_0 - 1/v_1)L \tag{4}$$

If $\Delta v_s < \Delta v_0$, the influence of linewidth can be ignored. $\Delta \lambda_0 = \lambda_s^2 \Delta v_0 / c$ as a function of the fiber length has been calculated in Fig. 4. It shows that $\Delta \lambda_0$ increases with core diameter while decreases with fiber length and core *NA*, which means the range that linewidth has negligible effect is larger for fiber with larger core, shorter length and smaller core *NA*. For the fiber with core *NA* smaller than 0.07 and fiber length shorter than 20m, the influence of linewidth can be neglected for linewidth smaller than 1nm.

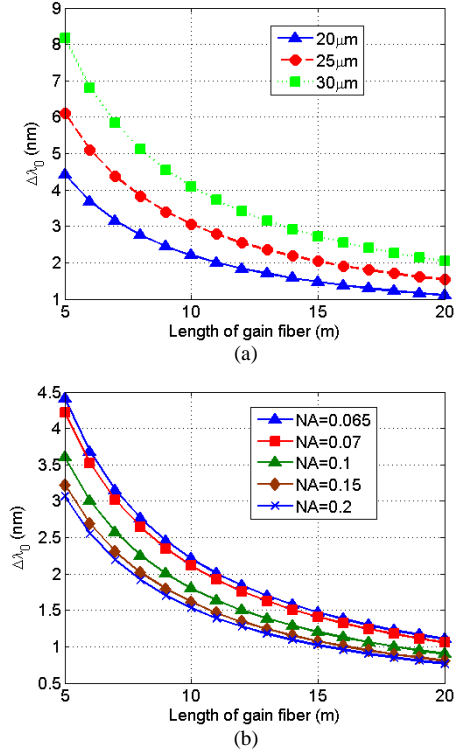

Fig. 4 $\Delta \lambda_0$ as a function of fiber length.

### III. EXPERIMENTAL STUDY

To verify the theoretical study, experimental study of core *NA* on MI has been carried out. The experiment setup is shown in Fig. 5. A broadband seed @1080nm, operating at the output power of 10W with 3dB linewidth of 0.2nm, was used to seed the amplifier. The main amplifier employed 30/250 LMA ytterbium-doped fiber (YDF). The core *NA* of the fiber is about 0.064. Six multimode fiber pigtailed 975 nm laser diodes (LD) are used to pump the gain fiber through a (6+1)×1 signal/pump combiner. A length of matched passive fiber is spliced to the end of the LMA YDF for power delivery. The spliced region is covered with high-index gel, which acts as cladding mode striper (CMS) to strip the residual pump laser and cladding mode. The output end of the delivery fiber is angle cleaved at 8°. The output laser was collected by a power meter. MI was monitored by detecting the time fluctuation of scattering power with photo-detector (PD) [28].

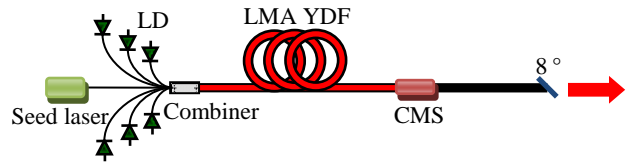

Fig. 5 Experimental setup of the high power fiber amplifier.

Typical results are presented in Fig. 6. Below the threshold, stable time traces with no fluctuation component was achieved, which correspond to stable beam profile. Above the threshold, the beam profiles became unstable with time traces fluctuating, which exhibits a periodic sawtooth-like oscillation. The threshold is stabilized around 376W after few tests including multiple power cycles.

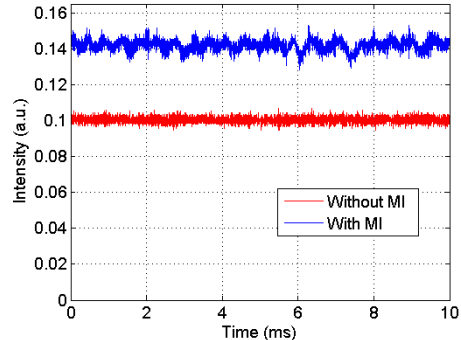

(a) Time traces

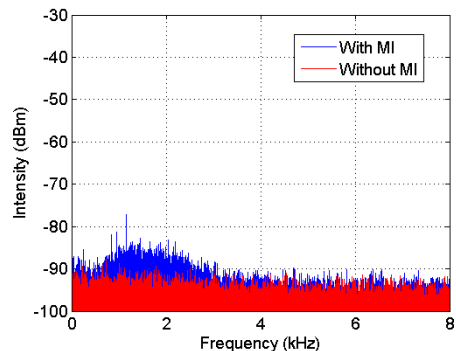

(b) Frequency distribution

Fig. 6 Typical time and frequency characteristics of MI.

Then the gain fiber of the main amplifier was replaced with 30/250 LMA fiber with core *NA* being 0.07. The threshold is measured to be stabilized around 367W and slightly lower than that for fiber with core *NA* being 0.064, which means that core *NA* has little impact on MI threshold for large core-cladding ratio and agrees with the aforementioned theoretical prediction on the impact of core *NA*. We also calculated the fraction of HOM as a function of laser power, which is shown in Fig. 7. The parameters are taken the same as in Table I except that the initial power and core *NA*, which are set to be the same as those in the experiment. The calculated threshold power is about 350W and 355W for *NA* being 0.07 and 0.064, respectively. The threshold agrees well with the experimental results, which means that the model is accurate for the case that the seed laser has a linewidth of 0.2nm and agrees with the aforementioned theoretical prediction on the effects of linewidth.

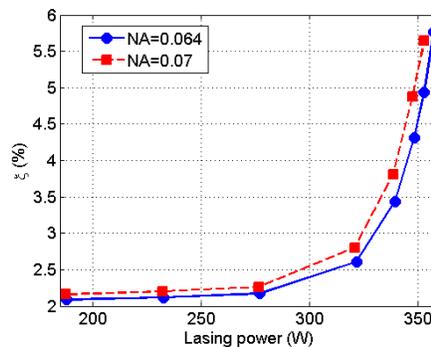

Fig. 7 Fraction of HOM as a function of output laser power.

## IV. Conclusions

In summary, we have investigated the effect of core *NA* and *V*-parameter on MI theoretically and experimentally. It shows that core *NA* has larger influence on MI for fibers with smaller core-cladding-ratio, and the influence of core *NA* on threshold is more obvious when pumped at 915nm. For the case that core *NA* decrease from 0.07 to 0.045, the threshold power increase by 57%, 25%, 16% and 11% for 20/400, 25/400, 30/400, 30/250 fiber, respectively. By comparing the results from aspect of *V*-parameter, it revealed that the threshold increases linearly as *V*-parameter decreases when *V*-parameter is larger than 3.5. When *V*-parameter is less than 3.5, the threshold shows exponentially increase as *V*-parameter decreases. We also discussed the effect of linewidth on MI, and found that the linewidth has negligible effect on MI for linewidth smaller than 1nm when the fiber core *NA* is smaller than 0.07 and fiber length is shorter than 20m. Fiber amplifiers with different core *NA* were experimentally analyzed and agreed with the theoretical predictions.

The authors would like to acknowledge the support of the National Science Foundation of China under grant No. 61322505, the program for New Century Excellent Talents in University.